# Superconductivity in FeTe$_{0.8}$S$_{0.2}$ induced by battery-like reaction

Aichi Yamashita[1,2], Satoshi Demura[1,2], Masashi Tanaka[1], Keita Deguchi[1,2], Takuma Yamaki[1,2], Hiroshi Hara[1,2], Kouji Suzuki[1,2], Yunchao Zhang[1,2], Saleem James Denholme[1], Hiroyuki Okazaki[1], Masaya Fujioka[1], Takahide Yamaguchi[1], Hiroyuki Takeya[1], Yoshihiko Takano[1,2]

[1]National Institute for Materials Science, 1-2-1 Sengen, Tsukuba, Ibaraki 305-0047, Japan

[2] Graduate School of Pure and Applied Sciences, University of Tsukuba, 1-1-1 Tennodai, Tsukuba 305-8571, Japan

**Abstract**

Superconductivity is successfully induced by utilizing a battery-like reaction found in a typical Li-ion battery. Excess Fe in FeTe$_{0.8}$S$_{0.2}$ is electrochemically de-intercalated by applying a voltage in a citric acid solution. The superconducting properties improve with an increase in the applied voltage up to 1.5 V. This result suggests that an electrochemical reaction can be used as a novel method to develop new superconducting materials.

**1. Introduction**

The iron-based superconductor, LaFeAsO$_{1-x}$F$_x$ was discovered by Kamihara *et al*. in 2008 [1]. This surprising news spread all over the world because given the typical ferromagnetic behavior of iron, it had been considered that it was incompatible with superconductivity. After the discovery, various kinds of iron-based superconductors have been found such as $Ln$FeAsO$_{1-x}$F (1111-system; $Ln$ = Lanthanoid elements), $Ae_{1-x}$K$_x$Fe$_2$As$_2$ (122-system; $Ae$ = Ca, Sr, Ba), $A$FeAs (111-system; $A$ = Li, Na), and so on. The superconducting transition temperature $T_c$ has been raised up to a maximum of 58 K in SmFeAsO$_{1-x}$F$_x$ to date [2]. The Iron-based superconductors have a layered structure which consists of an alternate stacking of blocking and superconducting layers.

A particular family of the iron chalcogenides superconductors (the 11-system), such as FeSe, Fe(Te,S), and Fe(Te,Se) [3, 4], have the simplest crystal structure among the iron-based superconductors. The iron chalcogenides are composed of only superconducting layers; in other words, there are no blocking layers in the structure. Therefore, investigation of the 11-system is important to understand the superconducting mechanism of the iron-based superconductors. In the 11-system, a small amount of Fe is trapped between the layers during the synthesis procedure. Excess Fe is difficult to remove by changing the stoichiometry in the conventional solid state synthesis. It is likely that excess Fe is normally incorporated into the structure of the 11-system. It is well known that excess Fe suppresses superconductivity [5]. Since the amount of excess Fe is considerable, it was reported that as-synthesized $FeTe_{0.8}S_{0.2}$ does not show superconductivity [6, 7].

Recently, we have been succeeded in inducing superconductivity in the 11-system compound by de-intercalating excess Fe from the interlayer by annealing in an organic acid solution [8, 9]. However, this reaction progresses slowly and is less tunable to induce superconductivity. Other method to de-intercalate excess Fe is required.

To find a new method, attention was focused on the chemical reaction which takes place in a typical Li-ion battery. During charge reaction the Li ion comes out from the positive electrode, namely de-intercalation, and then migrates to the negative electrode. In this study, the same principal is applied to the 11-system compound. Excess Fe in FeTe$_{0.8}$S$_{0.2}$ has been found to de-intercalate effectively and superconductivity has been induced.

**2. Experimental details**

Polycrystalline samples of FeTe$_{0.8}$S$_{0.2}$ were synthesized by a solid state reaction. A mixture of Te (Kojundo Chemical Lab. Co., Ltd., 99.999 %) and S (Kojundo Chemical Lab. Co., Ltd., 99.9999 %) grains, and Fe (Kojundo Chemical Lab. Co., Ltd., 99.9 %) powder with a nominal composition of FeTe$_{0.8}$S$_{0.2}$ was put into an evacuated quartz tube. After heating at 600 °C for 10 h, the obtained materials were thoroughly ground and pressed into disk-shaped pellets. The pellets were put into the evacuated quartz tube, and then reheated at 600 °C for a further 10 h. The electrochemical reaction was performed by the three electrodes method. Ag/AgCl was adopted as a reference electrode. Platinum plates were used for both working and counter electrodes. The sample was contacted with the working electrode by silver paste. Citric acid with a concentration of 6.0 (g/L) was used as a solution for the electrochemical reaction. And then a sample and the electrodes were soaked into a solution preheated at 80 °C. The voltage was applied for 1 hour at 1.0, 1.1, 1.2, 1.3, 1.4, 1.5, 1.7 V, respectively.

Powder X-ray diffraction (XRD) patterns were measured by a Rigaku model MiniFlex600 using Cu Kα radiation. The temperature dependence of magnetic susceptibility was measured using a SQUID magnetometer (Quantum design model MPMS) with both zero-field cooling (ZFC) and field cooling (FC) mode from 2.0 to 15.0 K under a magnetic field of 10 Oe. Fe and Te concentration in the solution after reaction was analyzed by using inductively-coupled plasma (ICP) spectroscopy (Thermo model iCAP-6200Duo).

**3. Results and discussion**

Figure 1 shows the X-ray diffraction patterns of $FeTe_{0.8}S_{0.2}$ samples of as-grown and the reacted samples. The obtained patterns are in good agreement with that of previous report. The peaks for the as-grown samples can be indexed using a tetragonal unit cell with lattice parameters $a$ = 3.811(2) Å, $c$ = 6.2412(2) Å. The X-ray diffraction patterns post reaction show no evidence of change the sample. And indexing of the structure produced almost the same lattice values ($a$ = 3.812(3) Å, $c$ = 6.241(2) Å).

Figure 2 shows the temperature dependence of magnetic susceptibility for the as-grown sample and the reacted samples. Since the as-grown sample is not a superconductor, it shows no diamagnetic signal. On the other hand, all reacted samples show a large diamagnetic signal at around 7 K in ZFC mode, which corresponds to superconductivity. This result indicates that superconductivity was successfully induced by the electrochemical reaction. The $T_c$ and the diamagnetic signal increased with increasing applied voltage up to 1.5 V. The shielding volume fraction was estimated from the value of ZFC mode at 2 K and plotted in figure 3 as a function of the applied voltage. The shielding volume fraction increased with increasing applied voltage, and showed a maximum at 1.5 V, and then drastically decreased at 1.7 V.

In order to elucidate what happened during the reaction process, each solution after the electrochemical reaction was analyzed using ICP spectroscopy. As a result, Fe and Te atoms were detected in the solution. The Fe and Te concentrations as a percentage were estimated from the detected amount of Fe and Te. The concentration was defined by the following formula,

$$\text{Fe or Te concentration (\%)} = \frac{\text{Detected Fe or Te in the solution (mol)}}{\text{Total Fe or Te of the sample before reaction (mol)}}$$

and then the concentration was plotted in figure 3. The Fe concentration in the solution increases with increasing applied voltage up to 1.5 V, while the Te concentration shows no significant increase. A strong correlation between the shielding volume fraction and the Fe concentration up to 1.5 V was observed. This result further indicates a clear correlation between the de-intercalation of excess Fe and the manifestation of superconductivity.

A small amount of the Te concentration in the solution suggests that the superconducting layer was not decomposed in the voltage range of 1.0-1.5 V. On the other hand, at 1.7 V, the Te concentration in the solution drastically increased to ~0.5 % suggesting that the superconducting layer has started to decompose around this voltage. Then the shielding volume fraction also decreased.

In conclusion, we have successfully induced superconductivity in $FeTe_{0.8}S_{0.2}$ by de-intercalation of excess Fe via an electrochemical reaction. The ICP spectroscopy analysis of the solution clearly showed that the Fe ion was effectively de-intercalated from the sample. The superconducting properties were improved with an increase of the applied voltage up to 1.5 V. This reaction is similar to the reaction in the Li-ion battery suggesting that the electrochemical reaction can be also applicable to other layered compounds for the development of new functional materials.

**Acknowledgments**

This work was partly supported by the Strategic International Collaborative Research

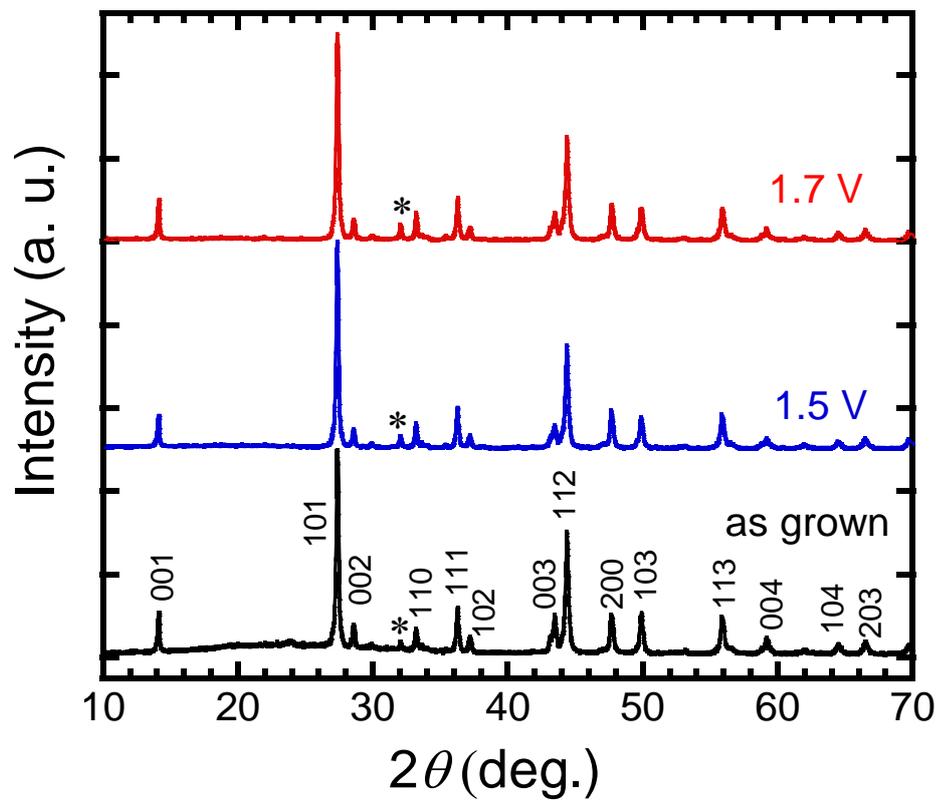

Figure 1.

XRD patterns for FeTe$_{0.8}$S$_{0.2}$ samples of the as-grown and after electrochemical reaction.

The asterisk in each pattern indicates a peak corresponds to an impurity phase of FeTe$_2$.

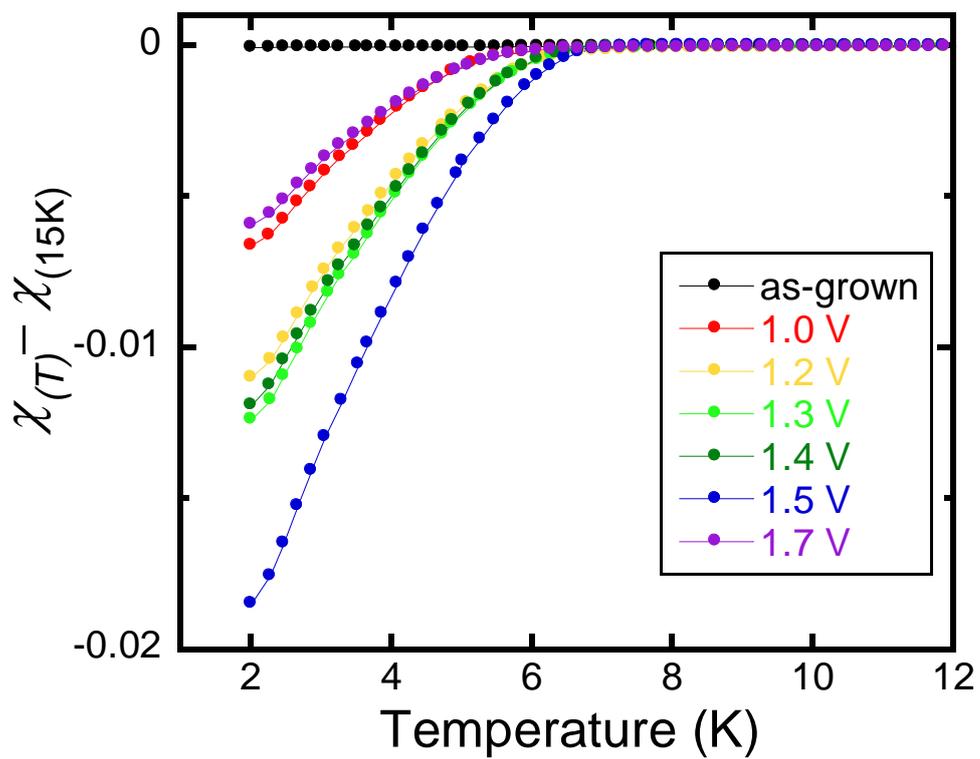

Figure 2.

Temperature dependence of magnetic susceptibility for the samples obtained by electrochemical reaction on various applied voltage. The susceptibility was subtracted at a value of 15 K.

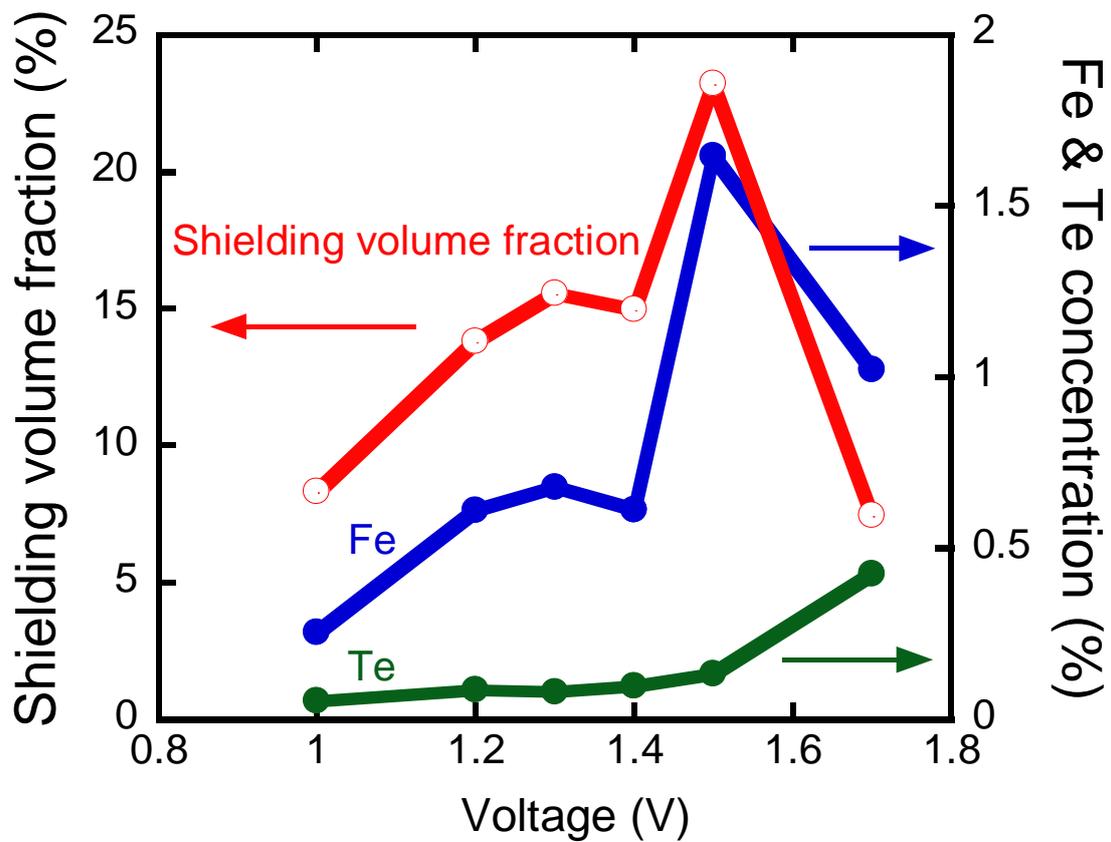

Figure 3.

Applied voltage dependence of shielding volume fraction (left hand side of the figure) and Fe, Te concentration in the solution after the electrochemical reaction (right hand side of the figure).